\documentclass[a4paper]{panl}
\usepackage{cite}
\usepackage{wrapfig}
\usepackage{graphicx}
\usepackage{amssymb}
\usepackage{amsfonts}
\usepackage{amsmath}
\usepackage{longtable}
\usepackage{rotating}
\usepackage{lscape}
\usepackage{epsfig}
\usepackage{multirow}

\originalTeX
\begin{document}
\def\dg{\dagger}

\title{Single excitation migration in molecular chain with an attached molecular structure: non-adiabatic polaron model}
\maketitle
\authors{D.\,Chevizovich$^{a}$\footnote{E-mail: cevizd@vin.bg.ac.rs},
S.\,Galovic$^{a}$,
V.\,Matic$^{a}$,
S.Eh.\,Shirmovsky$^{b}$,
D.V.\,Shulga$^{b}$, 
A.V.\,Chizhov$^{c}$
}
\setcounter{footnote}{0}
\from{$^{a}$\,``Vinca'' Institute of Nuclear Sciences, University of Belgrade, Serbia}
\from{$^{b}$\,Far Eastern Federal University, Vladivostok, Russia}
\from{$^{c}$\,Laboratory of Radiation Biology, Joint Institute for Nuclear Research, Dubna, Russia}

\begin{abstract}
В данной работе рассматривается возможность устойчивой миграции одиночного вибронного возбуждения в системе, состоящей из биомолекулы и присоединенной молекулярной структуры. Модель основана на предположении об автолокализации виброна и образовании неадиабатической поляронной квазичастицы. Показано, что, вопреки предсказанию неполяронной модели, возбуждение возникает на большом внутримолекулярном расстоянии от своего источника с высокой вероятностью.
\vspace{0.2cm}

In this paper we consider the possibility of the stable migration of the single vibron excitation in the system consisting of biomolecule and the attached molecular structure. The model is based on the assumption of the vibron self-trapping and the formation of the non-adiabatic polaron quasiparticle. We have shown that, contrary to the prediction of the non-polaron model, the excitation appears on a large intramolecular distance from its origin with a high probability.
\end{abstract}
\vspace*{6pt}


\label{sec:intro}
\section*{Introduction}

Processes in which different types of excitation migrate stably through biomolecular structures such as single charge injected in molecular chain (MC) or the quantum of energy created during the process of cellular respiration are of great importance for various physiological processes in the living cells~\cite{AK,DavydovBQM}. To explain the stability of these processes, various models have been developed. Some of them are based on the assumption that quantum effects play a key role here~\cite{AK,DavydovBQM,LBrizhikPRE2019, CDCSF2015,CDPRE2011}. Unfortunately, all of these models can not provide satisfactory answers to many arising questions. However, in the case of the migration of energy quanta through proteins, it seems that the model based on the assumption that this quanta becomes self-trapped (ST) forming so called non-adiabatic (small) polaron, provides the most adequate picture of the process of stable energy transfer along the biomolecular chain, over long distances~\cite{AK,DavydovBQM,CDPRE2011}.

There are many effects that significantly complicate understanding and theoretical modeling of energy quanta migration. Some of them are the influence of the environment in which the MC is placed and the existence of other molecular structures in the vicinity of the MC. The influence of the presence of additional molecules on the process of excitation ST has been poorly investigated so far. There are some studies related to adiabatic limit~\cite{LBrizhikPRE2019}, while non-adiabatic limit has remained almost unexplored. The stability of a possibly formed small polaron (SP) state in MC that is in contact with an attached molecular structure (D/A) has been examined in~\cite{CDPRE2024}. It was shown that the presence of the D/A molecule will not destroy the formation of the SP in the MC and for certain values of the system parameters the polaron state represents the favorable state of the excitation.

Based on the results of the paper~\cite{CDPRE2024}, here we study the problem of the migration of a single-vibron excitation along the MC in the presence of D/A molecule. The model was based on the assupmtion that the vibron, due to the excitation--phonon interaction, forms SP state. In the paper, we consider the possibility that the vibron initially located on the D/A molecule, over time can be found on a structural element (SE) of the MC, which is at large distance from the D/A molecule. We considered that this process occurs at the temperatures where living cells function. The obtained results are compared with the results predicted by the model where the excitation does not form a polaron state.

\section{The model}

Let us consider a quasi one-dimensional MC, whose SE interacts (and can exchange excitation) only with its nearest neighbors. Adjacent to some $m$-th SE of the MC there is an additional molecule D/A. It can exchange the excitation with the nearest SE from MC, but it does not affect the phonon spectrum of the MC itself (see Fig.~1).

\begin{figure}[h]
\begin{center}
\includegraphics[width=80mm]{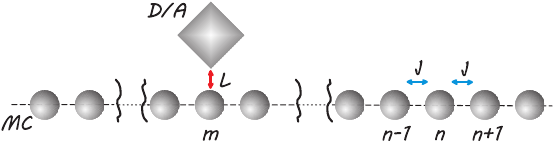}
\vspace{-3mm}
\caption{Schematic presentation of the considered MC--D/A structure.}
\end{center}
\labelf{fig01}
\vspace{-5mm}
\end{figure}

To theoretically describe the process of vibron ST in MC--D/A structure, we apply Holstein's model of MC, modified to take into account the presence of D/A molecule. At the initial moment, the vibron excitation is located on D/A molecule. After vibron deexcitation to the MC, due to the interaction with the phonons of the MC, vibron becomes self-trapped. Under the conditions defined in~\cite{CDPRE2024}, vibron forms stable small polaron state. Now, let us study the excitation migration through the MC--D/A structure, we start with the polaron Hamiltonian in the $n$-representation ~\cite{CDPRE2024}:

\begin{align}\label{Eq1}
\hat{\mathcal{H}}_{exc}&=
\mathcal{E}_D\hat{D}^{\dag}\hat{D}+\left(\mathcal{E}_0-\mathcal{E}_b\right)\sum_n\hat{B}^{\dag}_n\hat{B}_n-J\mathrm{e}^{-W_J(T)}\sum_n\hat{B}^{\dag}_n\left(\hat{B}_{n+1}+\hat{B}_{n-1}\right)\nonumber\\
&+L\mathrm{e}^{-W_L(T)}\left(\hat{B}^{\dag}_m\hat{D}+\hat{D}^{\dag}\hat{B}_m\right).
\end{align}

\noindent Here, $\hat{B}^{\dg}_n$ and $\hat{B}_n$ are creation and annihilation operators of the excitation on the $n$-th structure element of the MC. The $\mathcal{E}_0$ is the energy required to excite the corresponding excitation mode on the particular SE of MC, while the $\mathcal{E}_b$ is SP binding energy~\cite{CDPRE2024}. $J$ is the transfer integral between neighboring SE of MC. In the vibron case, it is the energy of the resonant dipole-dipole interaction. The $L$ is the energy of resonant dipole-dipole interaction between the attached molecule and the nearest molecular group of MC. Usually, it can be assumed that $|L|\le |J|$. Operators $\hat{D}^{\dg}$ and $\hat{D}$ are creation and annihilation operators of the excitation on the attached molecule with energy $\mathcal{E}_D$. Finally, $W_J(T)$ and $W_L(T)$ are the renormalization factors of $J$ and $L$, respectively~\cite{CDPRE2024}. 

To analyze the excitation appearance of on different SE of the MC, let us suppose that excitation at the initial moment $t=0$ is located on the attached molecule: $\left|\psi_i(0)\right\rangle=\hat{D}^{\dg}\left|0\right\rangle$. If there are no external influences, this state propagates in time as follows: $\left|\psi_i(t)\right\rangle=\mathrm{e}^{-\frac{i}{\hbar}\hat{\mathcal{H}}_{exc}t}\hat{D}^{\dg}\left|0\right\rangle$. The probability that, preforming the measure in the moment $t\neq 0$, we detect excitation at the $n$-th node of the molecular chain, i.e. in the state $\left|\psi_f(t)\right\rangle=\hat{B}^{\dg}_n\left|0(t)\right\rangle$, where $\left|0(t)\right\rangle=\mathrm{e}^{-\frac{i}{\hbar}\hat{\mathcal{H}}_{exc}t}\left|0\right\rangle$ is $v_{D,n}(t)=|\left\langle\psi_f(t)|\psi_i(t)\right\rangle|^2$. The probability amplitude of this event is $V_{D,n}(t)=\left\langle\psi_f(t)|\psi_i(t)\right\rangle=\left\langle 0\right|\mathrm{e}^{\frac{i}{\hbar}\hat{\mathcal{H}}_{exc}t}\hat{B}_n\mathrm{e}^{-\frac{i}{\hbar}\hat{\mathcal{H}}_{exc}t}\hat{D}^{\dg}\left|0\right\rangle$. In the Heisenberg representation it is:

\begin{equation}\label{VDnt}
V_{D,n}(t)=\left\langle 0\right|\hat{B}_n(t)\hat{D}^{\dg}(0)\left|0\right\rangle
\end{equation}

\noindent The further procedure is based on solving the system of equations of motion for amplitudes $V_{D,n}(t)$, $\forall n=1,2,...,N$. For the linear molecular structure with $N$ nodes and the attached molecule to the $m$-th SE of the MC, we have:

\begin{align}\label{dVlin}
i\hbar\dot{V}_D&=\mathcal{E}_DV_D+\bar{L}V_m\nonumber\\
i\hbar\dot{V}_1&=(\mathcal{E}_0-\mathcal{E}_b)V_1-\bar{J}V_2\nonumber\\
&.................................................................\nonumber\\
i\hbar\dot{V}_n&=(\mathcal{E}_0-\mathcal{E}_b)V_n-\bar{J}(V_{n-1}+V_{n+1});\;
n\neq m\nonumber\\
&.................................................................\\
i\hbar\dot{V}_m&=(\mathcal{E}_0-\mathcal{E}_b)V_m-\bar{J}(V_{m-1}+V_{m+1})
+\bar{L}V_D\nonumber\\
&.................................................................\nonumber\\
i\hbar\dot{V}_N&=(\mathcal{E}_0-\mathcal{E}_b)V_N-\bar{J}V_{N-1}\nonumber
\end{align}

\noindent here $\bar{J}=J\mathrm{e}^{-W_J(T)}$ and $\bar{L}=L\mathrm{e}^{-W_L(T)}$ 
are renormalized $J$ and $L$.

\section{Results and conclusions}

The distribution of $v_{D,n}(t)$ for the MC with $N=11$ nodes is presented in Fig.~\ref{fig02}. Here the normalized temperature is $k_BT/\hbar\omega_0=4$ (which correspond to room temperatures for chosen value of $\omega_0$), $\mathcal{E}_b/\hbar\omega_0=0.3$, $2J/\hbar\omega_0=0.1$, $|J|/|L|=0.1$ and $\mathcal{E_D}-\mathcal{E}_0=0.3\,\hbar\omega_0$. The time is presented in arbitrary units  $t\text{[a.u]}=t\text{[s]}\cdot\omega_0$, where $\omega_0=10^{13}\;\text{s}^{-1}$ \cite{AK,CDPRE2011,CDPRE2024,CDPB2016,CDCSF2015}.

\begin{figure}[h]
\begin{center}
\includegraphics[width=44mm]{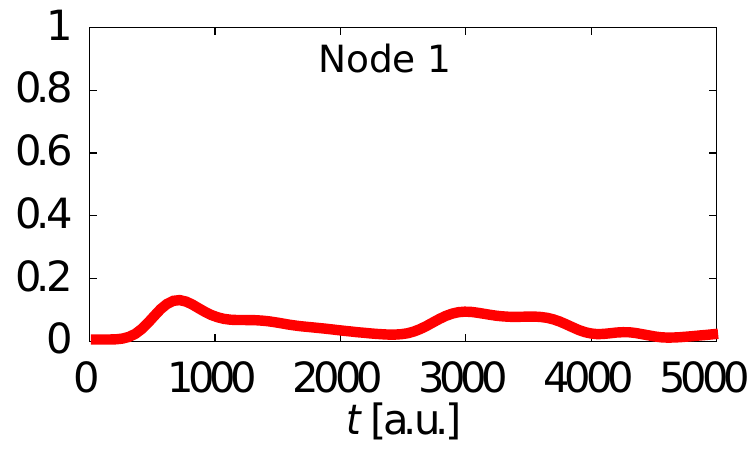}
\includegraphics[width=44mm]{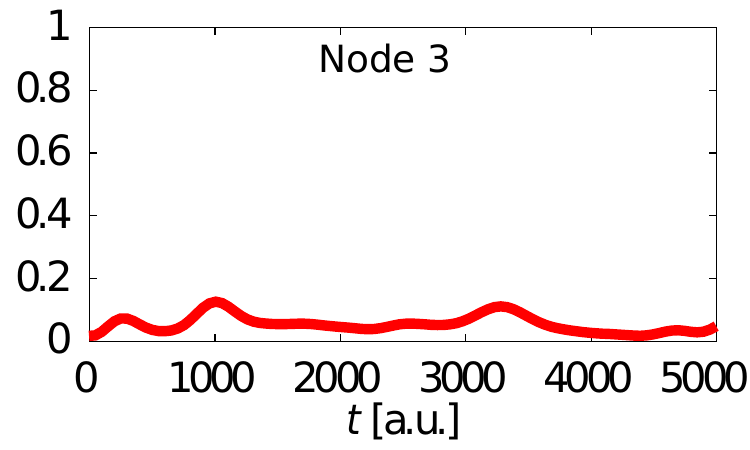}
\includegraphics[width=44mm]{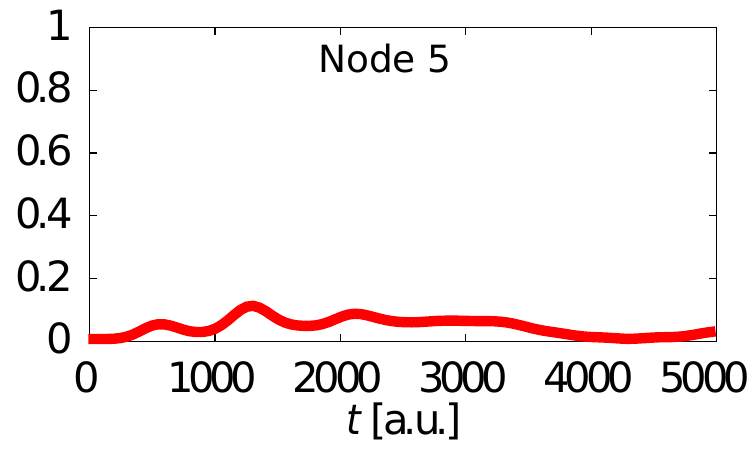}\\
\includegraphics[width=44mm]{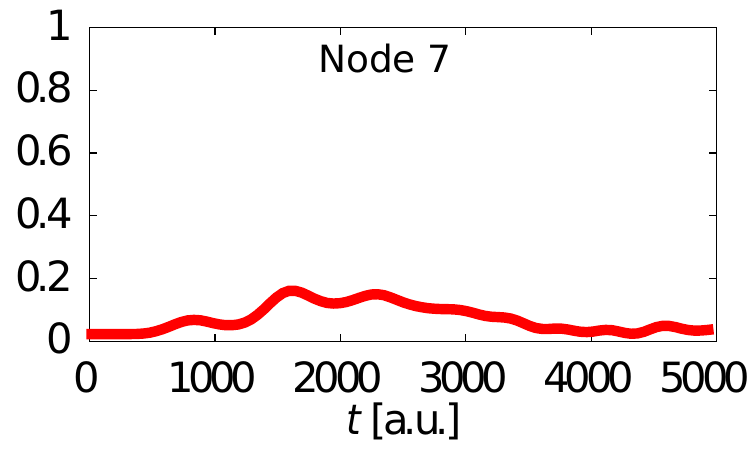}
\includegraphics[width=44mm]{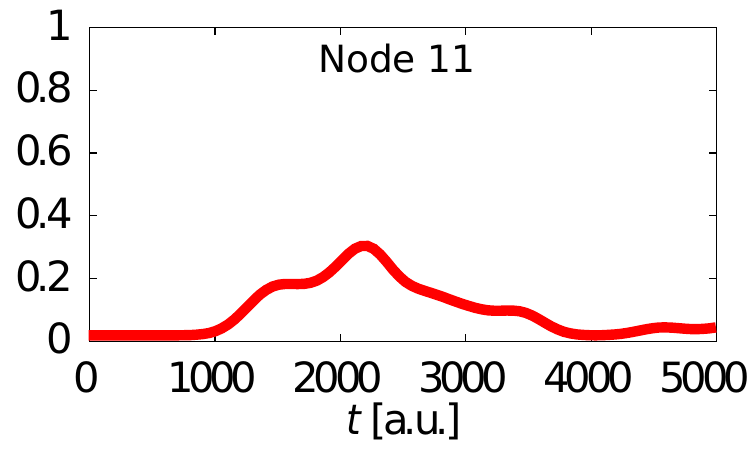}
\includegraphics[width=44mm]{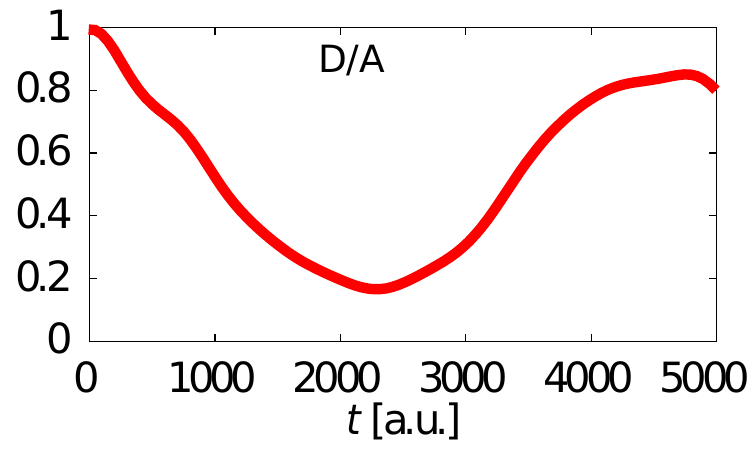}
\vspace{-3mm}
\caption{The probability distribution of the SP appearance on the different nodes 
of the MC during the time.}
\end{center}
\labelf{fig02}
\vspace{-5mm}
\end{figure}

According to the presented results, we can see that the probability of finding excitation at a large distance from the place of its origin (Node 11) can be significant (about 0.4) even at room temperatures. This occurs for the system parameter values that are characteristic for biomolecules. Such result points to a possible important role of the polaron effect in the processes of energy quantum transport during biophysical processes inside a living cell. It is interesting that the (significant) retention time of the excitation at the farthest node of the chain is of the order of a picosecond, which means that this excitation can affect the biochemical processes in which that part of the MC eventually participates.

Perhaps more interesting consequence of such a result would be in the case of an electron injected into a DNA, when molecule is exposed to the "soft" electromagnetic radiation. Such radiation cannot damage the molecule, but it can excite an electron from highest occupied molecular orbital into lowest unoccupied molecular orbital. By the self-trapping of this electron, an electron-polaron could be created, which could migrate on some distant nucleotide. If the retention time of such a polaron on the nucleotide was long enough, it could affect the functions of the nucleotide during the process of DNA replication or the creation of a protein. In other words, the polaron effect could be the source of "migrating mutations" in DNA.

Comparing the obtained results with those obtained using the model that does not take into account the polaron effect (Fig.~3.) \cite{BSShCh2023,BSSh2024}, we observe that the excitation migrates from th D/A molecule and after a certain time appears on the SE of the MC. As in the polaron model, the probability of finding excitation at a more distant node is larger than the probability of finding it at closer nodes. However, in contrast to the polaron model, the time that the excitation spends on the distant SE is significantly shorter compared to the time predicted by the polaron model.

\begin{figure}[h]
	\begin{center}
		\includegraphics[width=99mm]{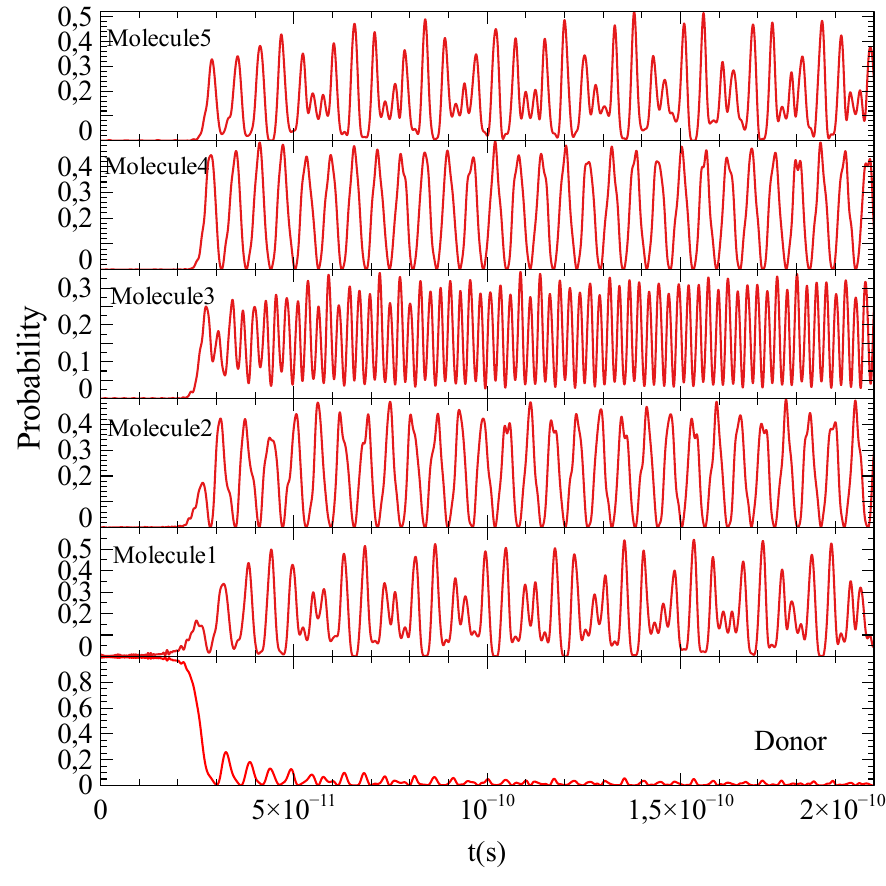}
		\vspace{-3mm}
		\caption{The probability distribution of excitation finding along the MC, obtained by the model that does not take into account the polaron effect.}
	\end{center}
	\labelf{fig03}
	\vspace{-5mm}
\end{figure}

Unfortunately, exact values of the system parameters of biomolecules do not belong to the strict non-adiabatic limit. Therefore, processes of energy migration in these systems should be investigated using one of the methods that interpolate between adiabatic and non-adiabatic limits. One of them is the polaron model based on the variational modification of the Lang--Firsov unitary transformation~\cite{CDPRE2011,CDPB2016}. Additional information on the efficiency of excitation transfer along a biomolecular chain, as well as the transfer of the excitation quantum state, can be obtained by studying the degree of correlation of quantum states of excitation at different nodes of the MC and possibly formed quantum entangled states of excitation between remote SEs~\cite{BSShCh2023,BSSh2024,NPCM2019,BioSys2024}.


\label{sec:funding}
\section*{Funding}
This work was partially supported by the Ministry of Science, Technological Development and Innovations of the Republic of Serbia (Contract No~451--03--47/2024--01/200017), the Projects within the Cooperation Agreement between the JINR, Dubna, Russian Federation and the Republic of Serbia (P21).



\begin{thebibliography}{1}
\def\selectlanguageifdefined#1{
\expandafter\ifx\csname date#1\endcsname\relax
\else\selectlanguage{#1}\fi}
\providecommand*{\href}[2]{{\small #2}}
\providecommand*{\url}[1]{{\small #1}}
\providecommand*{\BibUrl}[1]{\url{#1}}
\providecommand{\BibAnnote}[1]{}
\providecommand*{\BibEmph}[1]{\emph{#1}}
\ProvideTextCommandDefault{\cyrdash}{\hbox to.8em{--\hss--}}
\providecommand*{\BibDash}{\ifdim\lastskip>0pt\unskip\nobreak\hskip.2em\fi
\cyrdash\hskip.2em\ignorespaces}

\bibitem{AK}
\selectlanguageifdefined{english}
\BibEmph{Alexander D. M., Krumhansl J. A.} {Localized excitations in hydrogen-bonded molecular crystals}\href{https://doi.org/10.1103/PhysRevB.33.7172}{Phys. Rev. B}
\BibDash
\newblock 1986. \BibDash
\newblock V.~33 \BibDash
\newblock 7172

\bibitem{DavydovBQM}
\selectlanguageifdefined{russian}
\BibEmph{Давыдов А. С.} {Биология и квантновая механика}\BibDash
\newblock Наукова Думка, Киев, 1979.

\bibitem{LBrizhikPRE2019}
\selectlanguageifdefined{english}
\BibEmph{Brizhik L. S., Luo J., Piette B. M. A. G.,Zakrzewski W. J.} {Long--range donor--acceptor electron transport mediated by $\alpha$ helices}~//
\href{https://doi.org/10.1103/PhysRevE.100.062205}{Phys. Rev. E}
\BibDash
\newblock 2019. \BibDash
\newblock V.~100 \BibDash
\newblock 062205

\bibitem{CDCSF2015}
\selectlanguageifdefined{english}
\BibEmph{Cevizovic D., Ivic Z., Toprek D., Kapor D., Przulj Z.} {The influence of the interchain coupling on large acoustic polarons in coupled molecular chains: Three coplanar parallel molecular chains}~//
\href{https://doi.org/10.1016/j.chaos.2015.01.002}{Chaos, Solitons and Fractals}
\BibDash
\newblock 2015. \BibDash
\newblock V.~73 \BibDash
\newblock pp.71--79.

\bibitem{CDPRE2011}
\selectlanguageifdefined{english}
\BibEmph{Cevizovic D., Galovic S., Ivic Z.} {Nature of the vibron self-trapped states in hydrogen-bonded macromolecular chains}~//
\href{https://doi.org/10.1103/PhysRevE.84.011920}{Phys. Rev. E}
\BibDash
\newblock 2011. \BibDash
\newblock V.~84 \BibDash
\newblock 011920

\bibitem{CDPRE2024}
\selectlanguageifdefined{english}
\BibEmph{Matic V., Ivic Z., Przulj Z., and Chevizovich D.} {Influence of donor or acceptor presence on excitation states in molecular chains: Nonadiabatic polaron approach}~//
\href{https://doi.org/10.1103/PhysRevE.109.024401}{Phys. Rev. E}
\BibDash
\newblock 2024. \BibDash
\newblock V.~109 \BibDash
\newblock 024401

\bibitem{BSShCh2023}
\selectlanguageifdefined{english}
\BibEmph{Shirmovsky S. Eh., Chizhov A. V.} {Modeling of the entangled states transfer processes in microtubule tryptophan system}~//
\href{https://doi.org/10.1016/j.biosystems.2023.104967}{BioSystems}
\BibDash
\newblock 2023. \BibDash
\newblock V.~231 \BibDash
\newblock 104967

\bibitem{BSSh2024}
\selectlanguageifdefined{english}
\BibEmph{Shirmovsky S. Eh.} {Modeling of the quantum entangled state transfer protocol in the cell microtubules}~//
\href{https://doi.org/10.1016/j.biosystems.2023.105100}{BioSystems}
\BibDash
\newblock 2024. \BibDash
\newblock V.~235 \BibDash
\newblock 105100

\bibitem{CDPB2016}
\selectlanguageifdefined{english}
\BibEmph{Cevizovic D., Ivic Z., Galovic S., Reshetnyak A., Chizhov A. V.} {On the vibron nature in the system of two parallel macromolecular chains: The influence of interchain coupling}~//
\href{http://dx.doi.org/10.1016/j.physb.2016.02.033}{Phys. B}
\BibDash
\newblock 2016. \BibDash
\newblock V.~490 \BibDash
\newblock pp.9--15

\bibitem{NPCM2019}
\selectlanguageifdefined{english}
\BibEmph{Chizhov A. V., Chevizovich D., Ivic Z., Galovic S.} {Temperature dependence of quantum correlations in 1D macromolecular chains}~//
\href{DOI 10.17586/2220-8054-2019-10-2-141-146}{Nanosys.: Phys. Chem. Math.}
\BibDash
\newblock 2019. \BibDash
\newblock V.~10(2) \BibDash
\newblock pp.141--146

\bibitem{BioSys2024}
\selectlanguageifdefined{english}
\BibEmph{Shirmovsky S. Eh.} {On the possibility of implementing a quantum entanglement distribution in a biosystem: Microtubules}~//
\href{https://doi.org/10.1016/j.biosystems.2024.105320}{BioSystems}
\BibDash
\newblock 2024. \BibDash
\newblock V.~245 \BibDash
\newblock 105320

\end{thebibliography}

\end{document}